\input harvmac
%\draftmode
\let\includefigures=\iftrue
\let\useblackboard=\iftrue
\newfam\black

%Figure Stuff
\includefigures
\message{If you do not have epsf.tex (to include figures),}
\message{change the option at the top of the tex file.}
\input epsf
\def\figin{\epsfcheck\figin}\def\figins{\epsfcheck\figins}
\def\epsfcheck{\ifx\epsfbox\UnDeFiNeD
\message{(NO epsf.tex, FIGURES WILL BE IGNORED)}
\gdef\figin##1{\vskip2in}\gdef\figins##1{\hskip.5in}% blank space
instead
\else\message{(FIGURES WILL BE INCLUDED)}%
\gdef\figin##1{##1}\gdef\figins##1{##1}\fi}
\def\DefWarn#1{}
\def\figinsert{\goodbreak\midinsert}
\def\ifig#1#2#3{\DefWarn#1\xdef#1{fig.~\the\figno}
\writedef{#1\leftbracket fig.\noexpand~\the\figno}%
\figinsert\figin{\centerline{#3}}\medskip\centerline{\vbox{
\baselineskip12pt\advance\hsize by -1truein
\noindent\footnotefont{\bf Fig.~\the\figno:} #2}}
%\bigskip
\endinsert\global\advance\figno by1}
%%%
\else
\def\ifig#1#2#3{\xdef#1{fig.~\the\figno}
\writedef{#1\leftbracket fig.\noexpand~\the\figno}%
%\figinsert\figin{\centerline{#3}}\medskip
%\centerline{\vbox{\baselineskip12pt
%\advance\hsize by -1truein\noindent
%\footnotefont{\bf Fig.~\the\figno:} #2}}
%\bigskip\endinsert
\global\advance\figno by1} \fi

\def\id{{1 \kern-.28em {\rm l}}}

\def\K3{{\bf K3}}
\def\journal#1&#2(#3){\unskip, \sl #1\ \bf #2 \rm(19#3) }
\def\andjournal#1&#2(#3){\sl #1~\bf #2 \rm (19#3) }

\def\bar{\overline}

\def\ie{{\it i.e.}}
\def\eg{{\it e.g.}}

\def\tilde{\widetilde}

\def\frac#1#2{{#1\over#2}}

\def\half{\frac12}

\def\inbar{\,\vrule height1.5ex width.4pt depth0pt}
\def\IC{\relax\hbox{$\inbar\kern-.3em{\rm C}$}}
\def\IR{\relax{\rm I\kern-.18em R}}
\def\IP{\relax{\rm I\kern-.18em P}}

%
%%%%%%%%%%%%%%%%%%%%%%%%%%%%%%%%%%%%
%

%
\catcode`\@=11
\def\slash#1{\mathord{\mathpalette\c@ncel{#1}}}
\overfullrule=0pt

\def\SS{{\cal S}}

\def\underrel#1\over#2{\mathrel{\mathop{\kern\z@#1}\limits_{#2}}}

\catcode`\@=12

%%%%%%%%%%%%%%%%%%%%%%%%%%%%%%%%%%%%%%%%%%%%%%%%%%%%%%%%%%%%%%

%

\def\det{{\rm det}}

\def\det{{\rm det}}
\def\exp{{\rm exp}}

%%%%%%%%%%%%%%%%%%%%%%%%%%%%%%%%%%%%%%%%%%%%%%%%%%%%%%%%%%%%%%
% new defs:

\def\alphabar{{\bar\alpha}}
\def\betabar{{\bar\beta}}

\def\sigmabar{{\bar\sigma}}

%\NambuTP
\lref\NambuTP{Y.~Nambu and G.~Jona-Lasinio,
``Dynamical Model Of Elementary Particles Based On An Analogy With
Superconductivity. I,''Phys.\ Rev.\  {\bf 122}, 345 (1961).
%%CITATION = PHRVA,122,345;%%
}

%\HatsudaPI
\lref\HatsudaPI{
T.~Hatsuda and T.~Kunihiro,
``QCD phenomenology based on a chiral effective Lagrangian,''
Phys.\ Rept.\  {\bf 247}, 221 (1994)
[arXiv:hep-ph/9401310].
%%CITATION = HEP-PH 9401310;%%
}

%\VolkovKW
\lref\VolkovKW{
M.~K.~Volkov and A.~E.~Radzhabov,
``Forty-fifth anniversary of the Nambu-Jona-Lasinio model,''
arXiv:hep-ph/0508263.
%%CITATION = HEP-PH 0508263;%%
}

\lref\KlevanskyQE{
  S.~P.~Klevansky,
  ``The Nambu-Jona-Lasinio model of quantum chromodynamics,''
  Rev.\ Mod.\ Phys.\  {\bf 64}, 649 (1992).
  %%CITATION = RMPHA,64,649;%%
}

\lref\BuballaQV{
  M.~Buballa,
  ``NJL model analysis of quark matter at large density,''
  Phys.\ Rept.\  {\bf 407}, 205 (2005)
  [arXiv:hep-ph/0402234].
  %%CITATION = HEP-PH 0402234;%%
}

%\OsipovJS
\lref\OsipovJS{
A.~A.~Osipov, A.~E.~Radzhabov and M.~K.~Volkov,
``pi pi scattering in a nonlocal Nambu -- Jona-Lasinio model,''
arXiv:hep-ph/0603130.
%%CITATION = HEP-PH 0603130;%%
}

%\GrossJV
\lref\GrossJV{
D.~J.~Gross and A.~Neveu,
``Dynamical Symmetry Breaking In Asymptotically Free Field Theories,''
Phys.\ Rev.\ D {\bf 10}, 3235 (1974).
%%CITATION = PHRVA,D10,3235;%%
}

%\MosheXN
\lref\MosheXN{
M.~Moshe and J.~Zinn-Justin,
``Quantum field theory in the large N limit: A review,''
Phys.\ Rept.\  {\bf 385}, 69 (2003)
[arXiv:hep-th/0306133].
%%CITATION = HEP-TH 0306133;%%
}

%\GiveonSR
\lref\GiveonSR{
A.~Giveon and D.~Kutasov,
``Brane dynamics and gauge theory,''
Rev.\ Mod.\ Phys.\  {\bf 71}, 983 (1999)
[arXiv:hep-th/9802067].
%%CITATION = HEP-TH 9802067;%%
}

\lref\KarchSH{
  A.~Karch and E.~Katz,
  ``Adding flavor to AdS/CFT,''
  JHEP {\bf 0206}, 043 (2002)
  [arXiv:hep-th/0205236].
  %%CITATION = HEP-TH 0205236;%%
}

%\KruczenskiUQ
\lref\KruczenskiUQ{
  M.~Kruczenski, D.~Mateos, R.~C.~Myers and D.~J.~Winters,
  ``Towards a holographic dual of large-N(c) QCD,''
  JHEP {\bf 0405}, 041 (2004)
  [arXiv:hep-th/0311270].
  %%CITATION = HEP-TH 0311270;%%
}

%\BabingtonVM
\lref\BabingtonVM{
  J.~Babington, J.~Erdmenger, N.~J.~Evans, Z.~Guralnik and I.~Kirsch,
  ``Chiral symmetry breaking and pions in non-supersymmetric gauge /  gravity
  duals,''
  Phys.\ Rev.\ D {\bf 69}, 066007 (2004)
  [arXiv:hep-th/0306018].
  %%CITATION = HEP-TH 0306018;%%
}

%\SakaiCN
\lref\SakaiCN{
T.~Sakai and S.~Sugimoto,
``Low energy hadron physics in holographic QCD,''
Prog.\ Theor.\ Phys.\  {\bf 113}, 843 (2005)
[arXiv:hep-th/0412141].
%%CITATION = HEP-TH 0412141;%%
}

%\ItzhakiTU
\lref\ItzhakiTU{
N.~Itzhaki, D.~Kutasov and N.~Seiberg,
``I-brane dynamics,''
JHEP {\bf 0601}, 119 (2006)
[arXiv:hep-th/0508025].
%%CITATION = HEP-TH 0508025;%%
}

%\ItzhakiDD
\lref\ItzhakiDD{
N.~Itzhaki, J.~M.~Maldacena, J.~Sonnenschein and S.~Yankielowicz,
``Supergravity and the large N limit of theories with sixteen
supercharges,''
Phys.\ Rev.\ D {\bf 58}, 046004 (1998)
[arXiv:hep-th/9802042].
%%CITATION = HEP-TH 9802042;%%
}

%\WittenZW
\lref\WittenZW{
E.~Witten,
``Anti-de Sitter space, thermal phase transition, and confinement in  gauge
theories,''
Adv.\ Theor.\ Math.\ Phys.\  {\bf 2}, 505 (1998)
[arXiv:hep-th/9803131].
%%CITATION = HEP-TH 9803131;%%
}

%\KruczenskiUQ
\lref\KruczenskiUQ{
  M.~Kruczenski, D.~Mateos, R.~C.~Myers and D.~J.~Winters,
  ``Towards a holographic dual of large-N(c) QCD,''
  JHEP {\bf 0405}, 041 (2004)
  [arXiv:hep-th/0311270].
  %%CITATION = HEP-TH 0311270;%%
}

%\PolchinskiRR
\lref\PolchinskiRR{
J.~Polchinski,
``String theory. Vol. 2: Superstring theory and beyond,''
Cambridge University Press, 1998.
%\href{http://www.slac.stanford.edu/spires/find/hep/www?irn=4634802}{SPIRES entry}
}

%\PeskinEV
\lref\PeskinEV{
  M.~E.~Peskin and D.~V.~Schroeder,
 {\it An Introduction to quantum field theory,}
 Reading, USA: Addison-Wesley (1995).
 %\href{http://www.slac.stanford.edu/spires/find/hep/www?irn=3485960}{SPIRES entry}
}

\lref\LukyanovNJ{
S.~L.~Lukyanov, E.~S.~Vitchev and A.~B.~Zamolodchikov,
``Integrable model of boundary interaction: The paperclip,''
Nucl.\ Phys.\ B {\bf 683}, 423 (2004)
[arXiv:hep-th/0312168].
%%CITATION = HEP-TH 0312168;%%
}

%\KutasovRR
\lref\KutasovRR{
D.~Kutasov,
``Accelerating branes and the string / black hole transition,''
arXiv:hep-th/0509170.
%%CITATION = HEP-TH 0509170;%%
}

%\LukyanovBF
\lref\LukyanovBF{
S.~L.~Lukyanov and A.~B.~Zamolodchikov,
``Dual form of the paperclip model,''
arXiv:hep-th/0510145.
%%CITATION = HEP-TH 0510145;%%
}

%\MaldacenaIM
\lref\MaldacenaIM{
J.~M.~Maldacena,
``Wilson loops in large N field theories,''
Phys.\ Rev.\ Lett.\  {\bf 80}, 4859 (1998)
[arXiv:hep-th/9803002].
%%CITATION = HEP-TH 9803002;%%
}

%\ReyIK
\lref\ReyIK{
S.~J.~Rey and J.~T.~Yee,
``Macroscopic strings as heavy quarks in large N gauge theory and  anti-de
Sitter supergravity,''
Eur.\ Phys.\ J.\ C {\bf 22}, 379 (2001)
[arXiv:hep-th/9803001].
%%CITATION = HEP-TH 9803001;%%
}

%\BrandhuberER
\lref\BrandhuberER{
A.~Brandhuber, N.~Itzhaki, J.~Sonnenschein and S.~Yankielowicz,
``Wilson loops, confinement, and phase transitions in large N gauge  theories
from supergravity,''
JHEP {\bf 9806}, 001 (1998)
[arXiv:hep-th/9803263].
%%CITATION = HEP-TH 9803263;%%
}

%\'tHooftBH
\lref\tHooftBH{
G.~'t Hooft,
  ``Naturalness, Chiral Symmetry, And Spontaneous Chiral Symmetry Breaking,''
PRINT-80-0083 (UTRECHT)
%\href{http://www.slac.stanford.edu/spires/find/hep/www?r=print-80-0083\%2F(utrecht)}{SPIRES entry}
{\it Lecture given at Cargese Summer Inst., Cargese, France, Aug 26 - Sep 8, 1979}
}

%\SakaiYT
\lref\SakaiYT{
T.~Sakai and S.~Sugimoto,
``More on a holographic dual of QCD,''
Prog.\ Theor.\ Phys.\  {\bf 114}, 1083 (2006)
[arXiv:hep-th/0507073].
%%CITATION = HEP-TH 0507073;%%
}

%\tHooftHX
\lref\tHooftHX{
G.~'t Hooft,
``A Two-Dimensional Model For Mesons,''
Nucl.\ Phys.\ B {\bf 75}, 461 (1974).
%%CITATION = NUPHA,B75,461;%%
}

%%%%%%%%%%%%%%%%%%%%%%%%%%%%%%%%%%%%%%%%%%%%%%%%%%%
\Title{\vbox{\baselineskip12pt\hbox{EFI-06-05}
\hbox{}}} {\vbox{\centerline{NJL and QCD from String Theory}}}
\bigskip

\centerline{\it E. Antonyan, J.~A. Harvey, S. Jensen and D. Kutasov}
\bigskip
\centerline{EFI and Department of Physics, University of
Chicago}\centerline{5640 S. Ellis Av. Chicago, IL 60637}

\smallskip

\vglue .3cm

\bigskip

\bigskip
\noindent
We study a configuration of D-branes in string theory that
is described at low energies by a four-dimensional field theory
with a dynamically broken chiral symmetry. In a certain
region of the parameter space of the brane configuration
the low-energy theory is a non-local generalization of the
Nambu-Jona-Lasinio (NJL) model. This vector model is exactly
solvable at large $N_c$ and dynamically breaks chiral
symmetry at arbitrarily weak 't Hooft coupling. At strong
coupling the dynamics is determined by the low-energy theory
on D-branes living in the near-horizon geometry of other branes.
In a different region of parameter space the brane construction
gives rise to large $N_c$ QCD. Thus the D-brane system
interpolates between NJL and QCD.

\bigskip

\Date{March 2006}

\newsec{Introduction}

Quantum chromodynamics (QCD), the theory of the strong interactions,
is weakly coupled at high energies but strongly coupled at the scale
of typical hadron masses ($\sim 1$ GeV). It has proven difficult to
use this theory to study analytically the properties of low-lying
mesons and baryons.

In the approximation in which $N_f$ flavors of quarks are taken to
be massless, the Lagrangian of QCD has a chiral global symmetry
$U(N_f)_L\times U(N_f)_R$, acting on the left and right-handed quarks,
$q_L$, $q_R$. This symmetry is expected to be dynamically broken to the
diagonal subgroup $U(N_f)_{\rm diag}$. Analyzing this breaking is
difficult due to the strongly coupled nature of the theory.

An important early example of dynamical chiral symmetry breaking occurs in 
the Nambu-Jona-Lasinio (NJL) model \NambuTP. This model contains fermions 
$q_L$, $q_R$, which transform in the fundamental representation of a global
$U(N_c)$ symmetry, and interact via a local four Fermi coupling,
\eqn\lintnjl{\CL_{\rm int}=Gq_L^\dagger\cdot q_R q_R^\dagger\cdot q_L~,}
where the dot product denotes a contraction of the color indices. The
fermions $q_L$ and $q_R$ also transform in the fundamental representation
of $U(N_f)_L$ and $U(N_f)_R$ symmetries of $\CL_{\rm int}$, respectively.

The four Fermi interaction \lintnjl\ provides an attractive force between
the left and right-handed quarks. The idea of \NambuTP, based on an analogy
with the BCS treatment of superconductivity, was that this force may
destabilize the trivial vacuum and lead to the breaking of chiral symmetry.
To test whether this indeed occurs in the NJL model one must work with a
finite UV cut-off, since the interaction \lintnjl\ is non-renormalizable.
In \NambuTP\ it was shown that the resulting model breaks chiral symmetry
when the coupling $G$ exceeds a certain critical value.

The non-renormalizability of the NJL model implies that many of its predictions
are sensitive to the precise nature of the UV cut-off. Thus, it is natural
to look for a renormalizable field theory in which the ideas of \NambuTP\
can be tested in a more controlled setting. Such a setting is the two-dimensional
analog of the NJL model, which is asymptotically free. Moreover, since it is a
vector model\foot{See \eg\ \MosheXN\ for a review of large $N$ vector models.}
it is exactly solvable at large $N_c$ and finite $N_f$. This model was analyzed 
by D. Gross and A. Neveu \GrossJV, who showed that it breaks chiral symmetry
and generates a mass scale via dimensional transmutation.

The Gross-Neveu model provides a beautiful confirmation of the ideas of
Nambu and Jona-Lasinio, and makes it interesting to look for more
realistic four-dimensional models with similar properties. Restricting
to asymptotically free theories one is naturally led to QCD, since there
are no asymptotically free field theories in four dimensions without gauge
fields. Unfortunately QCD, like most models that involve $N_c\times N_c$
matrices, is difficult to solve even in the limit $N_c\to\infty$.

In this paper we explore a different route. We study a configuration
of D-branes in string theory that reduces at low energies to a field theory
of left and right-handed fermions $q_L$, $q_R$, with an adjustable attractive
interaction. The brane configuration we start with preserves a
$U(N_f)_L\times U(N_f)_R$ global symmetry but, as we will see, this
configuration (or vacuum) is unstable for all values of the coupling.
The true vacuum has a non-zero condensate $\langle q_L^\dagger\cdot q_R\rangle$.

At weak coupling, the infrared dynamics of $q_L$, $q_R$ is captured by a
non-local NJL model -- a vector model that does not contain dynamical gauge
fields and can be solved exactly in the limit $N_c\to\infty$, like the
Gross-Neveu model. It breaks chiral symmetry for arbitrarily weak coupling
and generates a mass for $q_L$, $q_R$. At strong coupling, the useful
description is in terms of D-brane dynamics in curved spacetime, which can
also be analyzed quite explicitly and exhibits similar properties.

The embedding of the NJL model in string theory that we describe provides
another bonus. Despite its non-renormalizability, the NJL model \lintnjl\
and its non-local generalizations have been used extensively in 
phenomenological studies of hadrons, and appear to give a rather accurate 
description of their properties (see \eg\ 
\refs{\KlevanskyQE\HatsudaPI\BuballaQV\VolkovKW-\OsipovJS} for reviews and
further references). However, it is not understood to what degree these models 
can be thought of as effective low-energy descriptions of QCD. The realization 
of NJL as a low-energy limit of the theory on D-branes provides a new perspective
on this problem. By varying the parameters of the D-brane configuration,
one can interpolate between NJL and QCD. It is likely that the interpolation
is smooth, \ie\ the two models are in the same universality class.

The idea that embedding field theories in string theory makes it easier to
understand otherwise mysterious phenomena in field theory is familiar from
other contexts. For example, thinking of $N=4$ SYM in four dimensions as
the low-energy limit of the six dimensional $(2,0)$ SCFT compactified on a
two-torus makes manifest the $SL(2,Z)$ S-duality symmetry of $N=4$ SYM,
which acts as the modular group of the torus. Thinking of $N=1,2$ SYM as
low-energy limits of the $(2,0)$ theory wrapped on other Riemann surfaces
provides a geometric realization of the Seiberg duality of $N=1$ SYM and of 
the Seiberg-Witten curves of $N=2$ SYM (see \eg\ \GiveonSR\ for a review).
Similarly, in our case both NJL and QCD are realized as low-energy limits of
the compactified $(2,0)$ theory in the presence of defects. It is interesting
that the $(2,0)$ theory plays a central role in all of these constructions.

The plan of the paper is the following. In section 2 we describe the brane
configuration and its massless excitations. We introduce the fundamental
length scales in the problem and discuss the range of validity of the weak
and strong-coupling approximations. In section 3 we study the low-energy
dynamics of the branes at weak coupling and show that it is described by a
non-local NJL model which exhibits chiral symmetry breaking at arbitrarily
weak coupling (for large $N_c$).

In section 4 we study the low-energy dynamics at strong coupling using a
description  in terms of D-branes in curved spacetime. We find the chiral
condensate and comment on the relation to the weak coupling analysis. In
section 5 we discuss a generalization of the D-brane construction which
interpolates between NJL and QCD. We conclude in section 6 with comments
on our results and a discussion of possible extensions of our analysis.
Our conventions and a few results used in the text are described in an
appendix.

\newsec{The D-brane configuration and some of its properties}

Motivated by earlier studies of string theory duals of QCD in the large
$N_c$ limit with the number of flavors $N_f$ held fixed (see e.g. 
\refs{\KarchSH\KruczenskiUQ\BabingtonVM-\SakaiCN}),
we will consider a brane configuration in $\IR^{9,1}$ that includes three
kinds of D-branes: $D4$, $D8$ and $\bar{D8}$. The different branes are
extended in the directions
\eqn\ddconfig{\eqalign{\qquad & 0 ~~~ 1 ~~~ 2 ~~~ 3 ~~~ 4 ~~~5
~~~ 6 ~~~ 7 ~~~ 8 ~~~ 9 ~~~ \cr
D4: ~~& {\rm x} ~~~ {\rm x} ~~~{\rm x} ~~~ {\rm x} ~~~ {\rm x}
~~~{} ~~~ {} ~~~{} ~~~ {} ~~~ {} ~~~ \cr
D8, ~ \bar{D8}: ~~& {\rm x} ~~~ {\rm x} ~~~ {\rm x} ~~~ {\rm x}
~~~ {} ~~~~ {\rm x} ~~~ {\rm x} ~~~ {\rm x} ~~~ {\rm x} ~~~ {\rm x} ~~~ \cr
}}
and are arranged\foot{This brane configuration was studied
in \SakaiCN, with the $x^4$ direction compact and with
anti-periodic boundary conditions for the fermions. As we will
see, the limit where the radius of $x^4$ goes to infinity is
particularly instructive. We will return to the compact case
in section 5.} as indicated in figure 1. Classically, the
$N_f$ $D8$ and $\bar{D8}$-branes are parallel and separated
by a distance $L$ in the $x^4$ direction. They intersect
the $N_c$ $D4$-branes at the origin of the $\IR^5$ labeled
by $(x^5, \cdots, x^9)$ and at $x^4=\pm {L/2}$. We will see
that this classical picture is modified by quantum effects.

\ifig\branecon{The configuration of $N_c$ $D4$-branes and $N_f$ $D8$ and
$\bar D8$-branes which leads to QCD in one limit and the NJL model in another.} 
{\epsfxsize3.0in\epsfbox{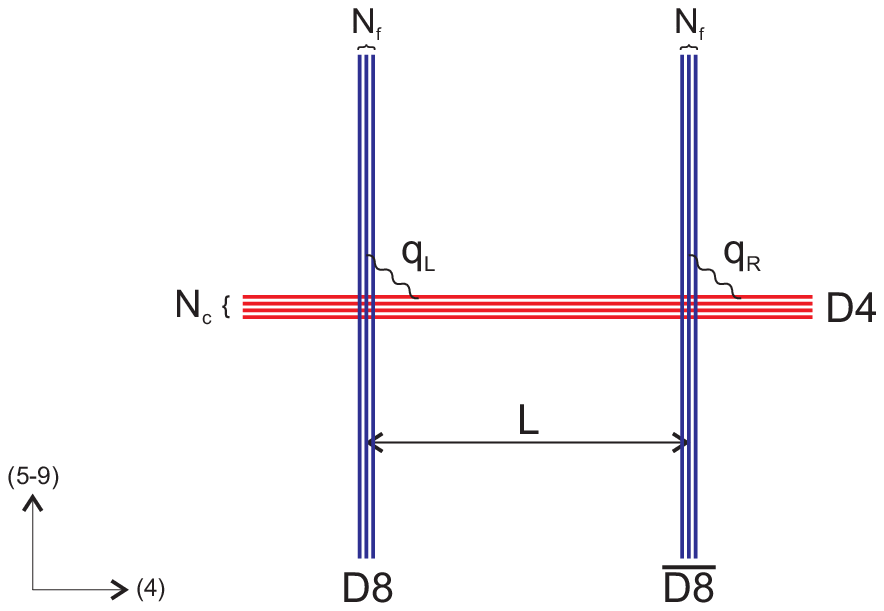}}

Since the different branes share the directions $(x^0,x^1,x^2,x^3)$,
it is natural to study the low-energy dynamics seen by a $3+1$-dimensional
observer living in the intersection region. In the
following sections we will do this in different regimes of the
parameter space of the brane configuration. Here, as preparation
for this study, we start with some general observations on this
problem.

As usual in D-brane physics, there are different sectors of
open strings that we need to consider. $p-p$ strings $(p=4,8,\bar 8)$
give rise to $p+1$-dimensional gauge theories with gauge groups
$U(N_c)$, $U(N_f)_L$ and $U(N_f)_R$ respectively. The field
content of these gauge theories is obtained by dimensional
reduction of $N=1$ SYM in ten dimensions. The gauge couplings
are (see \eg\ \PolchinskiRR)
\eqn\gsym{g_{p+1}^2=(2\pi)^{p-2}g_sl_s^{p-3}~.}
Since these theories live in more than $3+1$ dimensions they are
not of direct interest to us, as we are looking for modes localized
in the extra dimensions. The
\eqn\globsym{U(N_c)\times U(N_f)_L\times U(N_f)_R}
gauge symmetry of the D-branes gives rise to a global symmetry
of the $3+1$-dimensional theory.

Normalizable modes in $3+1$ dimensions are found in the $4-8$
and $4-\bar 8$ sectors, which contain states localized near
the corresponding brane intersection. Open strings in these
sectors have six Dirichlet-Neumann directions. The NS sector
states are massive; the only massless modes are spacetime
fermions. $4-8$ strings give a left-handed Weyl fermion $q_L$
localized at $x^4\simeq -L/2$, which transforms in the
$(N_c,N_f,1)$ of the global symmetry \globsym, while $4-\bar 8$
strings give a right-handed fermion localized at $x^4\simeq L/2$,
which transforms as $(N_c,1,N_f)$. A nice feature of the brane 
setup of \SakaiCN\ is that the left and right-handed fermions are 
separated in the extra dimensions.

The low-energy dynamics of the $D4-D8-\bar{D8}$ system can be
formulated in terms of $q_L$ and $q_R$. Alternatively, one can
integrate out these fields and study the dynamics of the gauge
fields in the presence of a localized source (see \ItzhakiTU\
for a related recent discussion). We will follow the former
approach, which is closer in spirit to the standard field
theoretic treatment of $3+1$-dimensional dynamics.

Consider first the case of a single intersection, of $N_c$
$D4$-branes and $N_f$ $D8$-branes, which can be thought
of as corresponding to the limit $L\to \infty$ of the
configuration of figure 1. We would like to study this
system in the limit $g_s\to 0$, $N_c \to \infty$ with
$g_sN_c$ and $N_f$ held fixed, and at energies much below
the string scale. In this limit the only interactions between
the fermions $q_L$ are due to their non-derivative couplings
to the $D4$-brane gauge fields. The leading effect of these
couplings is a four-Fermi interaction whose strength is
proportional to $g_5^2$, and whose precise form will be
described in the next section.

While the fermions $q_L$ live in $3+1$ dimensions, the gauge fields
that they exchange are $4+1$-dimensional. The `t Hooft coupling of
the five-dimensional gauge theory
\eqn\deflambda{\lambda={{g_5^2 \over 4 \pi^2} N_c~,}}
has units of length. This length determines the range of the
four-Fermi interaction. For example, one can show
that this interaction gives rise to corrections to the propagator
$\langle q_L^\dagger(x) q_L(y)\rangle$, which are suppressed relative
to the leading, free field, result  by powers of $\lambda/|x-y|$. Thus,
at distance scales much larger than $\lambda$ these interactions can
be neglected, while in the opposite regime $|x-y|\ll\lambda$ they are
large. We will discuss both limits below.

We next turn to the case where both the $D8$ and $\bar{D8}$-branes
are present. Since we would like to study the low-energy theory on
the branes, we will take the ratio of the distance between the
eight-branes and the string length, $L/l_s$, to be large but finite
in the limit $g_s\to 0$. This will allow us to neglect effects such
as the non-trivial dilaton and RR ten-form field strength created
by the $D8$-branes.

For finite $L$ there are interactions between the left-handed and
right-handed quarks which are due to exchange of $D4$-brane gauge
bosons. These interactions are weak for small $\lambda/L$, and
become stronger as one increases this parameter. As in other brane
systems, it is useful to label the different regimes by the value of
the dimensionless parameter  $g_sN_c = \lambda/l_s$. For small
$g_s N_c$ the hierarchy of scales in the problem is the following
\eqn\hiersc{\lambda\ll l_s\ll L~.}
As mentioned above, we will only be interested in the physics at 
distance scales much larger than $l_s$ (but which may be larger or 
smaller than $L$). In this regime we can neglect stringy effects on 
the dynamics of the fermions, as well as most of the effects of the
dynamics of the gauge fields on the $D4$ and $D8$-branes. The only 
effect we need to keep is the one gluon exchange between the left 
and right-handed quarks, since it gives a force between them which 
is absent at infinite $L$. Unlike the propagator corrections mentioned 
above, this is not a small correction to an existing effect, but rather 
the leading interaction in this channel.\foot{One can ask whether we 
should include  additional interactions between $q_L$ and $q_R$ due to 
exchange of the scalars on the $D4$-branes. These interactions provide 
small corrections to those due to gluon exchange, since the scalars are 
derivatively coupled to the quarks. For the scalars $\Phi_5,\cdots,\Phi_9$ 
this is due to the exact symmetry of the brane system 
$\Phi_i\to\Phi_i+{\rm const}$. $\Phi_4$ is a component of the $4+1$ 
dimensional gauge field, and is derivatively coupled to the quarks due to 
$4+1$ dimensional gauge invariance.}

Thus, in the limit \hiersc\ the dynamics is described by a field theory
of the fermions $q_L$ and $q_R$, with a non-local interaction due
to one (five-dimensional) gluon exchange. The theory contains a natural
UV cut-off $l_s$. Since this is much larger than $\lambda$, the
scale at which the non-linear dynamics of the five-dimensional gauge
field becomes important, we do not have to include
corrections to the Lagrangian obtained in the one gluon exchange
approximation. In the next section we will study the resulting non-local
NJL model, and show that its long distance dynamics is non-trivial.

As $g_sN_c$ increases, we get to the regime $l_s\ll\lambda\ll L$, in 
which we can still treat $q_L$ and $q_R$ as weakly interacting via 
single gluon exchange, but the natural UV cut-off is now $\lambda$ and 
not $l_s$. Further increasing $g_sN_c$ we get to $\lambda\sim L$, 
where multi-gluon exchange processes can no longer be neglected, and
the single gluon exchange approximation breaks down. For
\eqn\sscc{l_s\ll L \ll\lambda~,}
five-dimensional gauge theory effects give rise to a strongly
attractive interaction between $q_L$, $q_R$. This interaction can be
studied using the results of \ItzhakiDD, by analyzing the dynamics of
$D8$-branes in the near-horizon geometry of the $D4$-branes. This will
be discussed in section 4.

\newsec{Weak coupling analysis}

In this section we study the $D4-D8-\bar{D8}$ system in the weakly coupled
regime \hiersc. We start by deriving the low-energy effective action for the
fermions $q_L$, $q_R$. Consider first the case of a single intersection, say
that of $N_c$ $D4$-branes and $N_f$ $D8$-branes. Taking the intersection to 
be at $x^4=0$, the effective action for the Weyl fermion $q_L$ and $U(N_c)$ 
gauge field $A_M$ is given by
\eqn\lfoureight{\SS=\int d^5x \left[ -{1\over 4g_5^2}F_{MN}^2+
\delta(x^4) q_L^\dagger\bar\sigma^\mu(i\partial_\mu+A_\mu)q_L \right] ~.}
Note that while the first term (the gauge field Lagrangian) is integrated
over the $4+1$-dimensional worldvolume of the $D4$-branes with coordinates
$x^M$, $M=0,1,2,3,4$, the fermion Lagrangian is restricted to the 
$3+1$-dimensional intersection at $x^4=0$ with coordinates $x^\mu$, 
$\mu = 0,1,2,3$.

Since the physical degree of freedom at the intersection is the fermion
$q_L$, it is useful to integrate out the five-dimensional gauge field $A_M$.
Keeping only the quadratic terms in the gauge field Lagrangian, and
working in Feynman gauge, we find the following effective action for $q_L$: 
\eqn\fermeff{\SS_{\rm eff}=i\int d^4x q_L^\dagger\bar\sigma^\mu\partial_\mu q_L
-{g_5^2 \over 16 \pi^2} \int d^4x d^4y G(x-y,0)\left[ q^\dagger_L(x)\bar\sigma^\mu q_L(y)\right]
\left[q^\dagger_L(y)\bar\sigma_\mu q_L(x)\right]}
where $G(x^\mu,x^4)$ is proportional to the scalar propagator in $4+1$ dimensions,
\eqn\ggoo{G(x^\mu,x^4)={1\over  ( (x^4)^2-x_\mu x^\mu)^{3\over2}}~,}
The color indices in \fermeff\ are contracted in each term in brackets
separately, while the flavor ones are contracted between $q_L$ from one
term and $q_L^\dagger$ from the other.  Thus the quantities in square
brackets are color singlets and transform in the adjoint representation 
of $U(N_f)_L$. To derive \fermeff\ we used a Fierz identity described in 
appendix A.

The action \fermeff\ can be treated using standard large $N$ techniques. 
The solution has the following structure. For distance scales much larger
than $\lambda$ \deflambda, the field $q_L$ is free. The effects of the
interaction grow as the distance scale decreases, and become important
at distances of order $\lambda$.

At these distances there are two other types of interactions that are not
taken into account in \fermeff. One is due to the non-linear terms in the
gauge field Lagrangian \lfoureight. Their contributions are not small at
distances of order $\lambda$ and adding them brings back the full complexity
of the large $N_c$ gauge theory.
The other is due to string corrections. If the `t Hooft parameter $g_sN_c$
is small, the scale $\lambda$ \deflambda\ is much smaller than the string
scale (see \hiersc). Thus, before we get to the point where the gauge theory
effects become large, we reach a regime where the dynamics of $q_L$ is
dominated by exchange of massive string states. In this section we will
restrict the discussion to distances much larger than $l_s$ and $\lambda$,
for which we can neglect all these effects, such that $q_L$ is free in this limit.

We are now ready to discuss the case of interest, which contains
$D8$ and $\bar{D8}$-branes separated by a distance $L$. The analog
of the Lagrangian \lfoureight\ takes in this case the form
\eqn\llfour{\SS=\int d^5x \left[ -{1\over 4g_5^2}F_{MN}^2+
\delta(x^4+{L\over2}) q_L^\dagger\bar\sigma^\mu(i\partial_\mu+A_\mu)q_L+
\delta(x^4-{L\over2}) q_R^\dagger\sigma^\mu(i\partial_\mu+A_\mu)q_R \right] ~.}
Integrating out the gauge field in the single gluon exchange
approximation and neglecting interactions of the form in \fermeff\  we get
\eqn\lreff{\eqalign{
\SS_{\rm eff}= & i\int d^4x \left(q_L^\dagger\bar\sigma^\mu\partial_\mu q_L
+q_R^\dagger\sigma^\mu\partial_\mu q_R\right) \cr
& ~~~ +{g_5^2 \over 4 \pi^2} \int d^4x d^4y G(x-y,L) \left[ q^\dagger_L(x)\cdot q_R(y) \right]
\left[ q^\dagger_R(y)\cdot q_L(x) \right] \cr
}}
where we again used a Fierz identity from appendix A. As mentioned
above, \lreff\ provides an accurate description of the dynamics
of $q_L$, $q_R$ when $L$ and all other distance scales in the problem
are much greater than $\lambda$, $l_s$.

To solve the non-local Nambu-Jona-Lasinio model \lreff\ at large $N_c$
it is convenient to introduce a complex scalar field $T(x,y)$ which
transforms as $(1,\bar N_f, N_f)$ under the global symmetry \globsym,
and rewrite the quartic interaction
\lreff\ as follows:
\eqn\tlreff{\eqalign{
\SS_{\rm eff} =& i\int d^4x \left(q_L^\dagger\bar\sigma^\mu\partial_\mu q_L
+q_R^\dagger\sigma^\mu\partial_\mu q_R\right)\cr
&+\int d^4x d^4y \left[-{N_c\over\lambda}{T(x,y)\bar T(y,x)\over G(x-y,L)}+
\bar T(y,x)q^\dagger_L(x)\cdot q_R(y)+T(x,y)q^\dagger_R(y)\cdot q_L(x)\right]~.
\cr}}
The equation of motion of $T$ is
\eqn\eomtt{T(x,y)={\lambda\over N_c}G(x-y,L)q_L^\dagger(x)\cdot q_R(y)~.}
Plugging it back into \tlreff\ (or, equivalently, integrating out $T$,
$\bar T$), one recovers the original action \lreff. As usual in large $N$
theories, we would like instead to integrate out the fermions and obtain
an effective action for the scalars, whose dynamics becomes classical
in the limit $N_c\to\infty$. Since we are mainly interested in
the question of chiral symmetry breaking, we would like to compute the
expectation value of $T(x,y)$ in the vacuum. Poincare symmetry implies
that the latter must be a function of $(x-y)^2$. Thus, we will make this
simplifying assumption and compute the effective action for $T(x,y)=T(|x-y|)$.

To leading order in the $1/N_c$ expansion, the only corrections to the
classical action of $T$, $\bar T$ come from one loop diagrams with an
arbitrary number of external $T$, $\bar T$ fields. Computing the one-loop
effective potential in the standard way, adding to it the classical term
from \tlreff\ and dropping an overall factor of $N_c$ and the volume of
spacetime leads to the effective potential
\eqn\gapa{
V_{\rm eff}= \int d^4 x T(x) \bar T(x) {(x^2+L^2)^{3/2} \over \lambda}
- \int {d^4 k \over (2 \pi)^4} \log\left(1 + {T(k) \bar T(k) \over k^2}\right) }
where $T(k)$ is the Fourier transform of $T(x)$ (see appendix A for
conventions).  In \gapa\ we have also Wick rotated to Euclidean spacetime.

The equation of motion for $T$ that follows from \gapa\ (the {\it gap equation}) is
\eqn\gapeq{
\int d^4x {(x^2+L^2)^{3\over2}\over\lambda}T(x)e^{-ik\cdot x}=
{T(k)\over k^2+T(k)\bar T(k)}~.}
The trivial solution of this equation, $T(x)=0$, corresponds to a vacuum 
with vanishing $\langle q_L^\dagger\cdot q_R\rangle$ and unbroken
chiral symmetry. We will see that this solution is unstable. The true
vacuum has non-zero $T$ and breaks chiral symmetry.

The potential $V_{\rm eff}$ \gapa\ contains a classical term, which is
positive and favors $T=\bar T=0$, and a negative one loop term, which
becomes larger (\ie\ more negative) as $|k|$ decreases. Thus, it is
natural to expect that any non-trivial solution of \gapeq\ will be 
dominated by the low momentum modes of $T$. 

To find such a solution it is useful to discuss separately two regimes.
The first is the linear regime, in which
\eqn\linreg{T(k)\bar T(k)\ll k^2~,}
so that we can expand the $\log$ in \gapa\ and keep only the leading
quadratic term and hence the linear term in the gap equation \gapeq.
In this regime \gapeq\ becomes
\eqn\gapc{\nabla^2 \left[{(x^2 + L^2)^{3/2} \over \lambda}T(x) \right]
+ T(x) = 0~,}
where $\nabla^2$ is the four-dimensional Euclidean Laplacian.

As mentioned above, due to the relation with the chiral condensate,
\eomtt, we expect $T$ to depend only on  $r=\sqrt{x^2}$. In terms of
\eqn\deffr{F(r)=(r^2+L^2)^{3\over2}T(r)={\lambda\over N_c}
\langle q_L^\dagger(x)\cdot q_R(0)\rangle~,}
equation \gapc\ takes the form
\eqn\gapd{F''(r) + {3 \over r} F'(r) + {\lambda F(r)
\over (r^2+L^2)^{3/2}} = 0~.}
We now argue that the solution of \gapd\  which is relevant in the linear
regime \linreg\ is simply $F(r)=C$ with $C$ a constant to be determined.
To see this, note that \gapd\ can be solved in closed form for $r\ll L$
in terms of the dimensionless coordinate  $\rho = r \sqrt{\lambda/L^3}$
and for $r\gg L$ in terms of the dimensionless coordinate $\sigma=r/\lambda$.
The solution for $r \ll L$ which is regular as $\rho\to 0$ is expressed
in terms of Bessel functions as
\eqn\gapg{F(\rho)  \sim {J_1(\rho)\over\rho}~,}
and approaches a constant for small $\rho$. The regime of validity of
\gapg, $r\ll L$, corresponds to $\rho\ll\sqrt{\lambda/L}\ll1$. Thus, to
leading order in $\lambda/L$, the only part of the solution \gapg\ we are
sensitive to is its value at $\rho=0$.

For $r \gg L$ the leading behavior of the solution at large $\sigma$ can also
be expressed in terms of Bessel functions as
\eqn\gaph{F(\sigma) \sim {Y_2(2/\sqrt{\sigma}) \over \sigma}~,}
and approaches a constant for large $\sigma$. Since $r\gg L$ implies
$\sigma \gg1$, we are again only sensitive to this constant.

Thus, $F(r)$ is constant both for $r\ll L$ and for $r\gg L$. One can
show that it does not exhibit any non-trivial variation for $r\sim L$,
\ie\ the two constants are the same (to leading order in $\lambda/L$). 
This is a consequence of the fact that the coefficient of $F(r)$ in \gapd\ 
is bounded from above by $\lambda/L^3$, which means that the scale 
of variation of the solution is $\sqrt{L^3/\lambda}$, a distance scale that 
is much larger than $L$ in the weak coupling regime. It can also be 
verified by a numerical solution of \gapd.

In terms of $T(r)$ \deffr, the solution is
\eqn\tlarge{T(r) = {C \over (r^2 + L^2)^{3/2}}}
with Fourier transform
\eqn\momkk{T(k)={4 \pi^2 C e^{-k L} \over k}~.}
Since equation \gapc\ is linear, the constant $C$ is not determined by
the linear analysis. We will determine it below by matching
to the non-linear regime.

As we anticipated from the form of the potential \gapa, the
vacuum expectation value of $T$ \momkk\ grows with decreasing momentum.
The condition \linreg, which is necessary for the validity of the
preceeding discussion, is bound to be violated for sufficiently low
momenta. We will see that this happens in the regime $kL\ll 1$. Thus,
in the transition region we can neglect the exponential in \momkk, and
find that the linearity assumption breaks down at $k\simeq \sqrt C$.

We saw before that in the linear regime \linreg\ the gap equation \gapeq\
reduces to \gapc. In the opposite, strongly non-linear regime,
\eqn\nonlin{T(k)\bar T(k)\gg k^2~,}
one instead finds
\eqn\stnonl{
{1\over \bar T(k)} = \int d^4x {(x^2+L^2)^{3\over2}\over\lambda}T(x)e^{-ik\cdot x}
~.}
The solution of \stnonl\ is
\eqn\solnonll{\eqalign{
T(x)=&\bar T(x)=A\delta^4(x)~,\cr
T(k)=&\bar T(k)=A~,\cr
}}
where the constant $A$ is determined by \stnonl,
\eqn\formaa{A=\sqrt{\lambda\over L^3}~.}
The $\delta$-function behavior of \solnonll\ means that
$T(r)$ goes rapidly to zero for large $r$.

We can now determine the constant $C$ by matching the linear
and non-linear regimes. Comparing \momkk\ to the second line
of \solnonll\ we see that the transition between the two
occurs around the momentum $k^*$ which satisfies
\eqn\ckk{{4 \pi^2 C\over k^*}\simeq\sqrt{\lambda\over L^3}~.}
At that point we expect the parameter
\eqn\matchpar{{T(k^*)\bar T(k^*)\over (k^*)^2}\simeq
{A^2\over (k^*)^2}\simeq 1~.}
This means that
\eqn\detcc{C\simeq \left(k^*\over2 \pi\right)^2 \simeq
{\lambda\over 4\pi^2L^3}~.}
We are thus led to the picture shown in figure 2. For momenta
$k\gg\sqrt{\lambda / L^3}$ or, equivalently, distances
$|x|\ll \sqrt{L^3 / \lambda}$, we are in the linear regime,
in which the gap equation takes the form \gapc, and its
solution is given by \tlarge, \momkk. On the other hand, for
momenta $k\ll\sqrt{\lambda / L^3}$ the system is in the
non-linear regime, where the gap equation is given by \stnonl\
and its solution by \solnonll. The momentum scale
\eqn\kkoo{k^*=\sqrt{\lambda\over L^3}=
{1\over L}\sqrt{\lambda\over L}}
at which the transition between the linear and non-linear
regimes takes place is very low: $k^*\ll 1/L$ for weak
coupling, $\lambda\ll L$. This provides an aposteriori justification
for setting $\exp(-k^*L)\simeq 1$ in the analysis above.

\ifig\tk{Behavior of $T(k)$ as a function of momentum.} {\epsfxsize3.0in\epsfbox{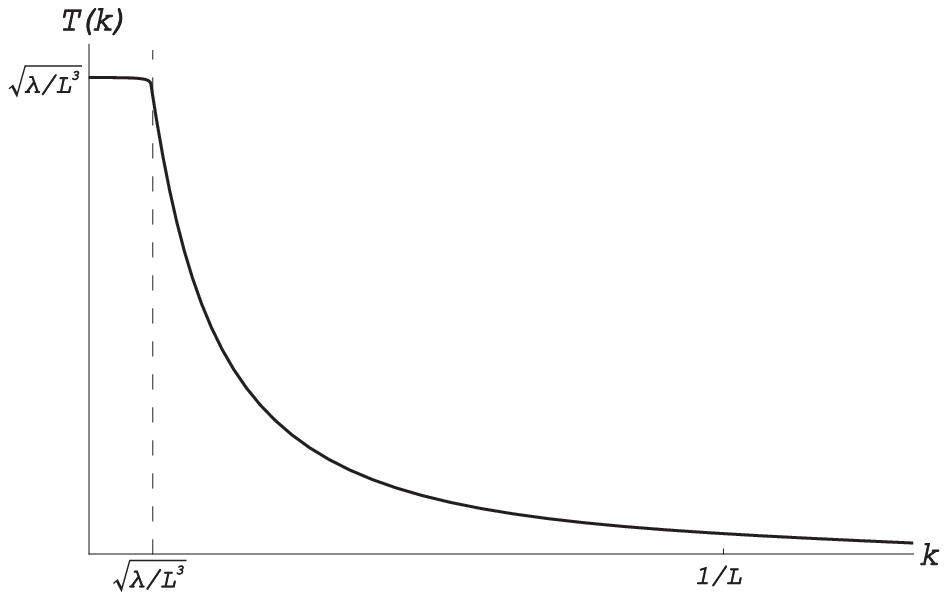 }}

To show that the non-local NJL model \lreff\ breaks chiral symmetry, it
is important to establish that the non-trivial solution of the gap 
equation constructed above has lower energy than the trivial solution 
$T=0$. This can be seen to be a property of any sufficiently well behaved
solution of the gap equation. 

Indeed, multiplying \gapeq\ by $e^{ik\cdot y}\bar T(y)$ and integrating 
over $k$ and $y$ leads to 
\eqn\neweq{{1\over\lambda}\int d^4x(x^2+L^2)^{3\over2}T(x)\bar T(x)
=\int{d^4k\over(2\pi)^4}{T(k)\bar T(k)\over k^2+T(k)\bar T(k)}~.}
Plugging this into \gapa\ we find that $V_{\rm eff}$ can be written
as
\eqn\newveff{V_{\rm eff}=\int{d^4k\over(2\pi)^4}
\left[{T(k)\bar T(k)\over k^2+T(k)\bar T(k)}-
\log\left(1+{T(k)\bar T(k)\over k^2}\right)\right]~.}
The integrand in \newveff\ is non-positive\foot{Indeed, defining 
$x=T(k)\bar T(k)/k^2$, the term in square brackets in  \newveff\
is given by $[\cdots]={x\over 1+x}-\log(1+x)$, which is negative for all 
$x>0$.} for any finite $T(k)$. Assuming that the integral over $k$ 
converges, which can be verified to be the case using the explicit 
form of $T(k)$ that we found, \momkk, \solnonll, one concludes 
that $V_{\rm eff}<0$ for the non-trivial solution of the gap equation, 
\ie\ the latter has lower energy than the trivial solution $T=0$.

We finish this section with a few comments:

\item{(1)} It is interesting to use the results above to calculate the quark
anti-quark condensate $\langle q_L^\dagger(x)\cdot q_R(y)\rangle$.
Using \deffr, \tlarge\ we see that in the linear regime
\eqn\qeqr{\langle q_L^\dagger(x)\cdot q_R(0)\rangle\simeq{N_c\over L^3}~.}
Thus, for $|x|=r\ll 1/k^*$ the chiral condensate is independent of $r$
and $\lambda$. For $r\gg 1/k^*$ it goes rapidly to zero. The full
solution can in principle be obtained numerically to arbitrarily high
precision.

\item{(2)} We emphasize again that the results obtained in this section are 
valid to leading order in $\lambda/L$. One can compute higher order corrections
to these results by including the non-linear terms in the gauge field
Lagrangian and other corrections discussed above.

\item{(3)} The chiral condensate \qeqr\ transforms in the $(\bar N_f, N_f)$ of 
the $U(N_f)_L\times U(N_f)_R$ global symmetry \globsym. The flavor indices
are suppressed above. The expectation value \qeqr\ breaks the chiral
symmetry $U(N_f)_L\times U(N_f)_R\to U(N_f)_{\rm diag}$.

\item{(4)} The condensate $\langle q_L^\dagger\cdot q_R\rangle$ that
we found in the non-local NJL model exhibits the type of behavior one
might expect in QCD. There, the chiral condensate should presumably be
roughly constant for $r$ smaller than $1/\Lambda_{QCD}$ and go to zero
for large $r$.

\newsec{Strong coupling analysis}

In the last section we analyzed the dynamics of the chiral fermions
$q_L$ and $q_R$ in the weakly coupled regime $\lambda\ll L$. As we
increase $g_sN_c$, or bring the $D8$-branes closer, the effective
coupling of the fermions due to exchange of $D4$-brane modes increases, 
and the approximations which we employed break down.

For $L\ll\lambda$, there is another weakly coupled description of the
dynamics. Instead of studying fermions coupled via
five-dimensional gluon exchange we should consider the dynamics of
$D8$-branes in the near-horizon geometry of the $N_c$ $D4$-branes.
In this section we will use this description to analyze the question
of chiral symmetry breaking in this regime.

Using the conventions of \ItzhakiDD\ in which the radial coordinate $U$
has dimensions of energy, the metric is given
by
\eqn\dfour{ds^2=\left( \alpha' U \over R\right)^{3\over2}
\left( \eta_{\mu \nu} dx^\mu dx^\nu-(d x^4)^2\right)-
\left(\alpha' U\over R\right)^{-{3\over2}}
\left((\alpha' dU)^2+(\alpha' U)^2d\Omega_4^2\right)~,}
where $\Omega_4$ labels the angular directions in $(56789)$.
The fourbrane geometry also has a non-trivial dilaton background,
\eqn\dfourdil{e^\Phi=g_s\left(\alpha' U\over R\right)^{3\over4}~.} %
The parameter $R$ is given by
\eqn\rdefn{R^3 =  \pi g_s N_c (\alpha')^{3/2} =
{g_5^2 \over 4 \pi} N_c \alpha' =  \pi \lambda \alpha'~.}
In this section we take $x^4$ to be non-compact. In the next section 
we will discuss the compact case studied in \SakaiCN. In what follows 
we set $\alpha'=1$.

We now consider a probe $D8$-brane propagating in the geometry
\dfour.\foot{The discussion of $N_f$ coincident $D8$-branes is 
essentially identical.} The $D8$-brane wraps $\IR^{3,1}\times S^4$ 
and forms a curve $U=U(x^4)$ in the $(U,x^4)$ plane, whose shape 
is determined by solving the equations of motion that follow from 
the DBI action on the $D8$-brane. In the background \dfour, \dfourdil\ 
the action is
\eqn\dbi{S_{D8} = -T_8V_{3+1}V_4 \int dx^4 U^4
\sqrt{1+ \left(R\over U\right)^3 U'^2}~,}
where $U'= dU/dx^4$, $V_{3+1}$ is the volume of $3+1$ dimensional
Minkowski spacetime, and $V_4$ the volume of a unit $S^4$. Since
the integrand of \dbi\ has no explicit $x^4$ dependence, $U(x^4)$
satisfies the first order equation \SakaiCN
\eqn\formdd{{U^4\over\sqrt{1+\left(R\over U\right)^3U'^2}}=U_0^4~.}
An interesting solution of \formdd\ is a U-shaped curve in the $(U,x^4)$ 
plane, which approaches $x^4=\pm {L\over2}$ as $U\to\infty$, and
is equal to $U_0$ at $x^4=0$, the point of closest approach of the
curve to $U=0$. The curve is symmetric under $x^4\to -x^4$.

Solving \formdd\ for $U'$ and integrating leads to
\eqn\tauint{x^4(U) = \int_{U_0}^U {dU \over \left( {U \over R}
\right)^{3/2} \left( {U^8 \over U_0^8} -1 \right)^{1/2} }}
which can be written in terms of complete and incomplete Beta functions
as
\eqn\tausoln{x^4(U) =  {1 \over 8} {R^{3/2} \over U_0^{1/2}}  \left[
B(9/16,1/2) - B(U_0^8/U^8; 9/16,1/2) \right] ~.}
{}From this we read off the asymptotic value
$x^4(\infty)=L/2$,
\eqn\taustar{L=
{1\over4}R^{3\over2}U_0^{-{1\over2}}B({9\over16},{1\over2})~.}
The incomplete Beta function has an expansion at small $z$ given by
\eqn\betaexp{B(z;a,b) = z^a \left[ {1 \over a} + {1-b \over a+1} z +
\cdots \right] ~.}
Keeping the first term in this expansion in \tausoln\ gives the form
of the curve at large $U$:
\eqn\asymshape{U^{9\over2}\simeq{2\over9}
{R^{3\over2}U_0^4\over {L\over2}- x^4}~.}
The part of the D-brane that corresponds to $x^4<0$ is determined
by the symmetry $U(x^4)=U(-x^4)$. The full $D8$-brane is shown
in fig. 3.

\ifig\dbsoln{A slice of the full $D8$-brane configuration showing the 
$D8$ and $\bar {D8}$-brane joined into a single $D8$-brane 
by a wormhole.} 
{\epsfxsize3.0in\epsfbox{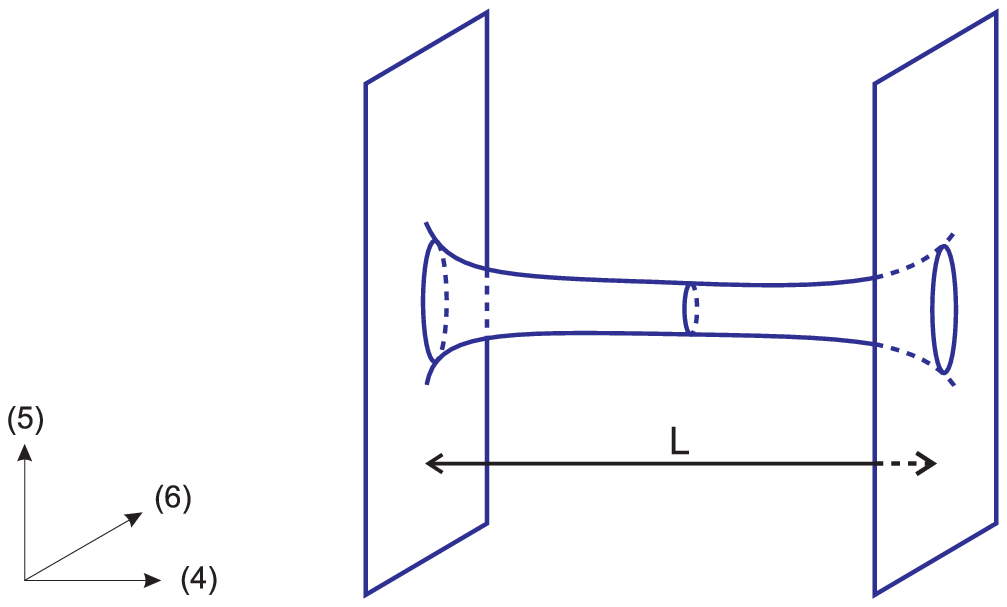}}

The solution described above is {\it not} the near-horizon
description of the brane configuration of figure 1. That
configuration corresponds in the geometry \dfour\ to a stack
of $N_f$ $D8$-branes stretched in $U$ and localized at
$x^4=-L/2$, and a stack of $N_f$ $\bar{D8}$-branes stretched
in $U$ and localized at $x^4=L/2$. In particular, unlike the
solution \tauint, the $D8$ and $\bar{D8}$-branes are not
connected in this case. Thus, the configuration of figure 1 preserves
a $U(N_f)_L\times U(N_f)_R$ symmetry acting separately
on the $D8$ and $\bar{D8}$-branes, unlike the configuration
\tauint\ where the two are connected and the symmetry is only
a single (diagonal) $U(N_f)$.

It is a dynamical question whether for fixed $L$ it is the separated
parallel brane configuration or the connected curved one, with $U_0$
a function of $L$ \taustar, that minimizes the energy density. The energy 
per unit volume is infinite, but the relative energy density of the two 
configurations is finite and can be computed by integrating the energy 
difference for each slice $dU$. After substituting the curved configuration's 
$U'$ from \formdd\ and rewriting the DBI action \dbi\ as an integral over 
$U$ rather than $x^4$, the difference 
$\Delta {E} \equiv {E}_{\rm straight}-{E}_{\rm curved}$ is proportional to
\eqn\endiff{\eqalign{
\Delta {E}  \sim &
  \int_0^{U_0} \left( U^{5/2} - 0 \right) dU
  + \int_{U_0}^\infty \left( U^{5/2}
    - U^{5/2} \left(1-\frac{U_0^8}{U^8}\right)^{-1/2} \right) dU
\cr
 =& -\frac{1}{8}U_0^{7/2} B(-7/16,1/2) \approx 0.052\, U_0^{7/2}
\;. \cr
}}
Thus, ${E}_{\rm straight} > {E}_{\rm curved}$,
so the curved configuration is preferred.  Physically, this is due to
the attractive force between the $D8$ and $\bar{D8}$-branes mediated
by the $D4$-brane fields. In the previous section we studied its
consequences in the weakly coupled regime. In the strongly coupled
regime under consideration here, the attractive force leads to a large
deformation of the $D8$-branes which can be seen in figure 3.

To check the validity of the supergravity approximation, it is
convenient to rewrite \taustar\ in terms of the variables $\lambda$
\deflambda\ and $L$. Omitting constants of order one, we have
\eqn\uuoo{U_0\simeq {\lambda\over L^2}~.}
As explained in \ItzhakiDD, a necessary condition for the supergravity
solution \dfour\ to be valid is that the curvature in string units is
small, which is the case when the effective `t Hooft coupling is large,
\eqn\effcoup{\lambda\left(U\over R\right)^{3\over4}\gg1~,}
or, equivalently,
\eqn\altuuoo{\lambda U\gg 1~.}
$U_0$ \uuoo\ satisfies \altuuoo\ if $\lambda\gg L$. Fixing $L$ and 
increasing $\lambda$ pushes $U_0$ further into the regime of validity of 
supergravity.\foot{There is an upper bound on $U_0$ coming from the 
requirement that $g_s(U_0)\ll1$, but it involves $N_c$ and we will not 
discuss it here.} On the other hand, decreasing $\lambda$ leads to 
smaller $U_0$, and eventually, when $\lambda\simeq L$ the curvature at 
$U_0$ becomes of order one and the supergravity description breaks down.

In section 3 we saw that the weakly coupled non-local NJL description
of the four-dimensional dynamics is valid for $\lambda\ll L$. Now we
see that the description in terms of a curved $D8$-brane is valid
for $\lambda\gg L$. This is an example of a bulk-boundary duality
analogous to that between $N=4$ SYM and supergravity in $AdS_5\times S^5$.
The analog of the `t Hooft coupling of $N=4$ SYM in our case is the
dimensionless ratio $\lambda/L$.

As emphasized in \SakaiCN, the fact that what looks asymptotically
like two disconnected stacks of $D8$ and $\bar{D8}$-branes is in fact
part of a connected stack of curved $D8$-branes provides a nice
geometric realization of chiral symmetry breaking. At high energies
(which correspond to large $U$ \ItzhakiDD) one sees an approximate
$U(N_f)_L \times U(N_f)_R$ symmetry while at low energies (small $U$)
only $U(N_f)_{\rm diag}$ is manifest.

The energy scale associated with chiral symmetry breaking in the 
supergravity regime is $U_0$ \uuoo. This scale can be thought of as the 
constituent quark mass of $q_L$, $q_R$. Indeed, in the limit $\lambda\gg L$ 
the spectrum contains free quark states which correspond to fundamental 
strings stretched between the curved $D8$-branes and $U=0$ (the location 
of the $N_c$ $D4$-branes). The energy of such strings is $U_0$. The analog 
of this scale at weak coupling is $k^*$ \kkoo. As in other bulk-boundary dualities, 
the mass goes like a different power of the coupling $\lambda/L$ in the weak 
and strong coupling regimes. 

To make the description of the $3+1$-dimensional dynamics
in terms of a curved $D8$-brane in the $D4$-brane geometry
more precise, we need to find the map between bulk fields
in the geometry \dfour\ and boundary operators. Once this
map is established, one can use the standard tools of
holographic dualities to study the boundary theory. In
particular, giving an expectation value to a non-normalizable
operator in the bulk corresponds to adding the dual boundary
operator to the Lagrangian. Normalizable bulk v.e.v.'s
correspond to giving expectation values to the dual boundary
operators.

In our system, there are two kinds of bulk fields. One is
closed string fields (such as the dilaton, graviton, etc)
in the geometry \dfour, which exist even when the $D8$-branes
are absent. The other is open string fields on the $D8$-branes.
Both live in the bulk (\ie\ at any $U$) and couple to $q_L$,
$q_R$. We will briefly comment on the bulk-boundary map for
some open string fields, leaving a more detailed analysis to
the future.

In order to find the boundary operators corresponding to different
open string modes we go back to the D-brane configuration we
started with (figure 1). Consider first the open string tachyon
stretched between the $D8$ and $\bar{D8}$-branes. This complex
scalar field which transforms as $(\bar N_f,N_f)$ under
$U(N_f)_L \times U(N_f)_R$ couples to the fermions via a
Yukawa-type interaction
\eqn\yuqq{\CL_1\simeq Tq_R^\dagger\cdot q_L+
\bar Tq_L^\dagger\cdot q_R~.}
Thus, in the near-horizon geometry of the $D4$-branes, the
open string tachyon $T$, $\bar T$ is dual to the boundary
operators
\eqn\tqeqr{\eqalign{
T&\leftrightarrow q_R^\dagger\cdot q_L~,\cr
\bar T&\leftrightarrow q_L^\dagger\cdot q_R~.\cr
}}
In particular, turning on a mass for $q_L$, $q_R$
corresponds to giving a non-normalizable expectation
value to the field $T$, while a v.e.v. for 
$q_R^\dagger\cdot q_L$ corresponds to a normalizable 
expectation value of $T$.

As we discussed above, the curved $D8$-brane \tauint\
describes a vacuum with non-zero expectation value
$\langle q_R^\dagger\cdot q_L\rangle$. Therefore, in
addition to its curved shape, the D-brane \tauint\ must
have a non-zero normalizable condensate of the $8-\bar 8$
tachyon $T$. At first sight this may seem odd, but in fact
something very similar is known to happen for the closely
related hairpin D-brane \refs{\LukyanovNJ\KutasovRR-\LukyanovBF}.
In our case, the fact that $T$ has an expectation value can be
seen from \yuqq. Since $\langle q_R^\dagger\cdot q_L\rangle$ is
non-zero, there is a tadpole for $T$ localized at the intersection.
Therefore, it has a non-zero expectation value, which is
also localized in the vicinity of $U=0$.

Another mode that we can consider is the scalar $\left(X^4\right)_L$
that parametrizes the location of the $D8$-branes in the $x^4$
direction, and its $\bar{D8}$-brane counterpart $\left(X^4\right)_R$.
Unlike the tachyon, $\left(X^4\right)_{L,R}$ do not have Yukawa-type
couplings to $q_L$, $q_R$, as can be easily deduced from symmetry
considerations. The lowest dimension coupling consistent with the
symmetries has the form 
\eqn\ltwoo{\CL_2\simeq \left(X^4\right)_L\left(\partial_\mu J^\mu_L\right)
+\left(X^4\right)_R\left(\partial_\mu J^\mu_R\right)~,}
where $J^\mu_L$, $J^\mu_R$ are the $U(N_f)_L\times U(N_f)_R$ currents.
In the free theory at infinite $L$ they are conserved, so the coupling
\ltwoo\ can be neglected. For finite $L$, chiral symmetry breaking 
implies that $\partial_\mu J^\mu_{L,R}\not=0$ and proportional to 
$q_L^\dagger\cdot q_R q_R^\dagger\cdot q_L$ (integrated over some
of the positions). Thus, \ltwoo\ can be written as
\eqn\ltwo{\CL_2\simeq \left(X^4\right)_L
\left(q_L^\dagger\cdot q_R q_R^\dagger\cdot q_L\right)_{(N_f^2,1)}
+\left(X^4\right)_R
\left(q_L^\dagger\cdot q_R q_R^\dagger\cdot q_L\right)_{(1,N_f^2)}
}
where the subscripts indicate that in the first term
$q_L^\dagger\cdot q_R q_R^\dagger\cdot q_L$ are coupled to an adjoint
of $U(N_f)_L$ and a singlet of $U(N_f)_R$, and similarly for the
second term.

In the near-horizon geometry \dfour, \ltwo\ implies that
the bulk-boundary correspondence is
\eqn\xfourlr{\eqalign{
\left(X^4\right)_L&\leftrightarrow
\left(q_L^\dagger\cdot q_R q_R^\dagger\cdot q_L\right)_{(N_f^2,1)}\cr
\left(X^4\right)_R&\leftrightarrow
\left(q_L^\dagger\cdot q_R q_R^\dagger\cdot q_L\right)_{(1,N_f^2)} ~. \cr
}}
The fact that the operators on the r.h.s. of \xfourlr\ have
non-zero expectation values in our brane configuration implies
that the scalars $\left(X^4\right)_L$, $\left(X^4\right)_R$
must have a non-zero normalizable expectation value. This is nothing
but the curved shape of the branes, which is given asymptotically
by \asymshape. It is normalizable since $x^4$ approaches $L/2$ as
$U\to\infty$ like $L/2-x^4\sim U^{-{9\over2}}$.

It would be interesting to study the correspondence between
bulk fields and boundary operators in more detail, but we will
not pursue that here. We finish this section with some comments 
on the $8-\bar 8$ string, which seems to play an important role 
in this system.

In flat space the ground state of the $8-\bar 8$ string is tachyonic for $L$ 
less than the string length and has positive mass squared for $L$ larger
than the string length (as is the case here). One might be tempted
to think that it can be decoupled from the dynamics because of its
string scale mass.  However we believe that this is incorrect.

One way to see this is to study the geometry of a fundamental string stretched 
between the $D8$ and $\bar{D8}$-branes in the near-horizon geometry of the 
$D4$-branes. Naively, such a string is stretched in the $x^4$ direction at fixed
$U$.  Let $U^*$ be the fixed value of $U$ and $L^*/2 = x^4(U^*)$. Then the 
coordinate length of the string is $L^*$, and including the warp factor in \dfour, 
its physical length is
\eqn\lstring{L(U)=L^*\left(U\over R\right)^{3\over4}~.}
However, this is not the lowest energy configuration. As is familiar
from the study of warped geometries \refs{\ReyIK\MaldacenaIM-\BrandhuberER},
the string dips into the small $U$ region due to the warping. To find
the amount of this dipping, we can proceed as follows.

The string is described by the Nambu-Goto action,
\eqn\sng{S_{NG}= - \int d^2\xi\sqrt{-\det h_{ab}}}
where
\eqn\indmetric{h_{ab}=\partial_a X^M\partial_b X^n G_{MN}}
is the induced metric on the string and we omitted an overall
numerical factor. Plugging \dfour\ into the Nambu-Goto action 
and omitting a factor of the length of time, we find that
\eqn\newacti{S_{NG}= - \int_{-{L^*\over2}}^{L^*\over2}d x^4
\sqrt{\left(U\over R\right)^3+U'^2} ~.}
The shape $U(x^4)$ satisfies the first order equation
\eqn\formfund{{\left(U\over R\right)^{3\over2}
\over\sqrt{1+\left(R\over U\right)^3U'^2}}=\left(U_0^{(F)}\over
R\right)^{3\over2}~.}
Here $U_0^{(F)}$ is the minimal value of $U$ that the fundamental
string attains,
at $x^4=0$, where $U'$ vanishes. It is determined by $U^*$ (and
$L^*$, but the latter is also determined by $U^*$ via the shape
of the $D8$-brane, which we take as given).

Some algebra leads to the following relation between the various
parameters:
\eqn\relfund{{L^*\over 2 U_0^{(F)}}=
\left(R\over U_0^{(F)}\right)^{3\over2}
\int_1^{U^*\over U_0^{(F)}}dx x^{-{3\over2}}(x^3-1)^{-\half} ~.}
This equation is valid for all $U^*$, but it simplifies for large
$U^*$,
for which $L^*$ approaches \taustar, and one can assume that
$U^*\gg U_0^{(F)}$ so that the upper limit of the integral \relfund\
can be
taken to infinity. This gives the following relation:
\eqn\asymrel{{L\over 2U_0^{(F)}}=
{1\over3}\left(R\over U_0^{(F)}\right)^{3\over2}B({2/3},{1/2})~.}
Note in particular that the amount by which the fundamental string
descends
into the $D4$-brane throat is independent of $U^*$ for large $U^*$.

We can also compare the minimal value of $U$ for the fundamental
string and the $D8$-brane. The latter is given by \taustar; combining
it with \asymrel\ we conclude that
\eqn\ratiouu{\left(U_0^{(F)}\over
U_0\right)^\half={8B({2/3},{1/2})
\over 3B({9/16},{1/2})}\simeq 2.38 ~.}
Thus, the fundamental string reaches the vicinity of the tip of the
$D8$-brane, but stays away from it by a finite amount. This behavior
is shown in figure 4.

\ifig\fstring{Configurations of fundamental strings with ends on the $D8$ and $\bar{D8}$-branes.} {\epsfxsize3.0in\epsfbox{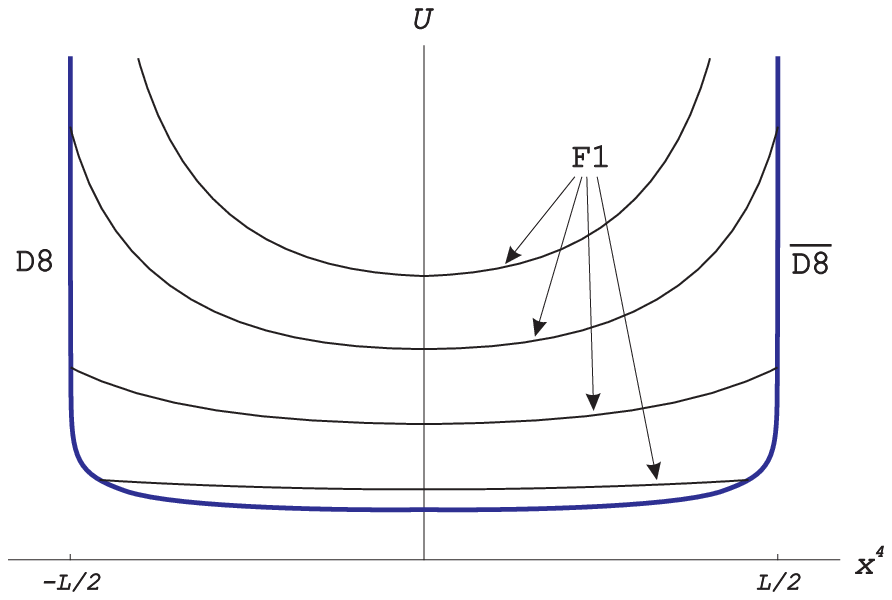}}

Another interesting quantity to compute is the length $L_{\rm curved}$ of the
fundamental string whose shape is described by \formfund. Dividing this length
by the length of the straight string (and omitting some numerical constants)
one finds that the ratio is
\eqn\ratioll{{L_{\rm curved}\over L_{\rm straight}}\sim \left(U^*\over
U_0^{(F)}\right)^{-{1\over2}} ~.}
Hence, for large $U^*$ the curved string is much shorter than the
straight one.

Thus, even a string that starts at a large value of $U$ and might be
expected to naively decouple from the infrared physics at small $U$
in fact does not. To minimize its length, the fundamental string
descends into the $D4$-brane throat,  reaching the vicinity
of the place where the $D8$-brane itself turns around, and
proceeds to the other side in $x^4$.

\newsec{NJL and QCD}

An interesting generalization of the brane configuration we have
been studying is obtained by compactifying the $x^4$ direction on
a circle of radius $R_4$, with antiperiodic boundary conditions for
the fermions living on the $D4$-branes. For finite $R_4$ the $U(N_c)$
gauge field on the $D4$-branes becomes dynamical. The antiperiodic
boundary conditions give a mass to the adjoint fermions and scalars
coming from $4-4$ strings, at tree level and one loop, respectively.
Below the mass of the adjoint fermions and scalars the dynamical
degrees of freedom are a $U(N_c)$ gauge field and $N_f$ fermions in
the fundamental representation of the gauge group.

This brane configuration was studied in \SakaiCN\ as a
description of QCD with $N_f$ flavors, generalizing the work of
\WittenZW, where the system without $D8$-branes (and thus
without fundamentals) was considered as a model for pure 
Yang-Mills theory. Its low-energy
dynamics depends on two dimensionless parameters: $\lambda/L$ and
$L/R_4$. In sections 2 -- 4 we considered the case $L/R_4=0$ and studied
the dependence of the dynamics on $\lambda/L$. In this section we
will qualitatively describe the dependence of the dynamics on $L/R_4$,
which varies over the range $[0,\pi]$. This will help us to relate
the discussion of sections 2 -- 4 of this paper to that of \SakaiCN,
and to QCD.

We start with the case where $x^4$ is compactified on a large
circle of radius $R_4\gg L$. As in the previous sections, the
dynamics depends on the open string coupling $g_s N_c$. For
weak coupling, $g_s N_c\ll 1$, the hierarchy of scales is
(compare to \hiersc)
\eqn\newhierwk{\lambda\ll l_s\ll L\ll R_4~.}
The fact that $R_4$ is finite means that the classical 
four-dimensional `t Hooft coupling $\lambda_4$
\eqn\lamfour{\lambda_4={\lambda\over 2\pi R_4}}
is finite as well. In the regime \newhierwk, the four-dimensional 
coupling $\lambda_4$ is small. Dimensional transmutation generates 
a dynamical scale $\Lambda_{QCD}$, at which four dimensional gauge 
interactions become large. 

For small $\lambda_4$, $\Lambda_{QCD}$ is much smaller 
than the dynamically generated mass of the quarks, $k^*$ \kkoo. 
Therefore, the dynamics splits into two essentially decoupled parts. 
The chiral symmetry breaking is still described by the non-local NJL 
model of section 3. The interactions of the quarks with four dimensional 
gauge fields provide a small correction to their properties at the scale 
$k^*$. In particular, $q_L$ and $q_R$ behave as free particles at distance 
scales of order $1/k^*$. The gauge fields introduce a confining potential 
for the quarks at the much larger distance scale $1/\Lambda_{QCD}$. 
Thus, for finite $R_4$ we expect to find a discrete spectrum of meson 
resonances with energies of order $2k^*$ and mass splittings of order 
$\Lambda_{QCD}$. 

As the open string coupling $g_sN_c$ increases, the five-dimensional 
coupling and $\Lambda_{QCD}$ increase as well. For
\eqn\intreg{l_s\ll L\ll\lambda\ll R_4}
the four-dimensional coupling \lamfour\ is still small, but
the five dimensional one is large. Thus, one can use the
supergravity description of section 4 (with small corrections
due to the finiteness of $R_4$) to describe the dynamical mass
generation of the quarks, while the confining interactions of
the quarks with the gauge field are still described by Yang-Mills
theory with a small `t Hooft coupling \lamfour. In particular, the
dynamics still splits into the chiral symmetry breaking part and
the confining part, which occur at different energy scales. 

Further increasing $g_sN_c$ we reach the regime
\eqn\largecoup{l_s\ll L\ll R_4\ll \lambda~.}
Here, both the four-dimensional and the five-dimensional
`t Hooft couplings are large, and one needs to use supergravity
to study both chiral symmetry breaking and confinement. The 
near-horizon metric of the $D4$-branes in this regime is given by
\eqn\fiveb{ds^2=\left(\alpha' U \over R\right)^{3\over2}
\left( \eta_{\mu \nu}dx^\mu dx^\nu- f(U)(d x^4)^2\right)-
\left(\alpha' U\over R\right)^{-{3\over2}}
\left({(\alpha' dU)^2 \over f(U)}+(\alpha' U)^2d\Omega_4^2\right)~,}
where
\eqn\fivec{f(U) = 1- \left(U_{KK} \over U\right)^3 ~,}
$R$ is given by \rdefn\
and
\eqn\ukk{U_{KK} = {4 \pi \over 9} {\lambda\over R_4^2}~.}
As is familiar from studies of near-extremal $D4$-branes, the
geometry \fiveb\ is smooth. The $U$ coordinate is restricted
to $U \ge U_{KK}$, such that the warp factor in $\IR^{3,1}$
as well as the volume of the four-spheres remain finite for all $U$.
When \largecoup\ is valid, both $U_0$ \uuoo, and $U_{KK}$ \ukk\
are in the region where the curvature of the metric \fiveb\ is
very small, so one can study low-lying open and closed strings using
supergravity.

Equation \largecoup\ implies that $U_0\gg U_{KK}$. Therefore,
the dynamics associated with chiral symmetry breaking is still
insensitive to the presence of dynamical four dimensional gauge
fields. Indeed, in studying the shape of the $D8$-branes in 
section 4 we took $f(U)=1$, rather than \fivec. However, this 
shape is only sensitive to $f(U\ge U_0)$. Thus, the analysis of 
section 4 is valid for finite $U_{KK}$, up to small corrections 
that go like $(U_{KK}/U_0)^3$. In particular, the dynamically
generated mass of the quarks is still given by $U_0$ \taustar, \uuoo.

What does change significantly in this case is the properties of
constituent quarks, which for $U_{KK}=0$ were described by
fundamental strings going from $U_0$ down to $U=0$. Indeed, 
consider a quark  anti-quark pair, which corresponds to a fundamental 
string whose ends lie on the $D8$-branes, and are separated by a 
distance $x$ in $\IR^3$. For $R_4\gg x\gg L$, this looks like a string going 
from the $D8$-brane down towards $U=0$ and another one with 
opposite orientation a distance $x$ apart. Its shape is insensitive to
the finite value of $R_4$. This implies that the quark and anti-quark  
are weakly interacting at these scales. 

When the distance $x$ becomes of order $R_4$, the string notices 
that it can descend no further in $U$ and its properties change. This 
corresponds to the confining potential created by the four dimensional 
dynamics at the distance scale $R_4$. Thus, in this regime the quarks
can be thought of as particles with the mass $U_0$ \uuoo\ which are 
free at short distances and form bound states whose size is of order 
$R_4$.

To summarize we find that if we compactify the system discussed in 
sections 2 -- 4 on a large circle of radius $R_4\gg L$, the chiral symmetry
breaking dynamics of the fermions $q_L$ and $q_R$ remains essentially 
unchanged and is almost decoupled from that of the four dimensional 
gluons of $U(N_c)$.  At the energy scales associated with chiral symmetry 
breaking  the four-dimensional coupling is very small and one can neglect 
the dynamics of the gauge fields. The scale $\Lambda_{QCD}$ at which the 
gauge coupling becomes strong is well below the dynamically generated
masses of the quarks. Gauge dynamics leads to the formation of bound 
states whose typical size is $1/\Lambda_{QCD}$.

To get a theory that looks more like QCD we need to take
the parameter $L/R_4$ to be of order one. For example, \SakaiCN\
discussed the case in which the $D8$ and $\bar{D8}$-branes are
maximally separated on the circle, which corresponds to $L=\pi R_4$.
When $L$ and $R_4$ are comparable, the only dimensionless parameter
that we can vary is $\lambda/L\sim \lambda/R_4$.

For $\lambda\ll L$, the low-energy theory on the branes is QCD with
$N_f$ massless fundamentals, and a small four-dimensional `t Hooft
coupling \lamfour. The QCD scale is well below the other scales, $1/L$,
$1/R_4$. This is the theory one would like to solve, but unfortunately,
there are no good analytical tools to study it in this regime.

For $\lambda\gg L$ the theory is not quite QCD, but it is expected to
be in the same universality class. In particular, qualitative phenomena
such as dynamical symmetry breaking, and the existence of a discrete
spectrum of massive glueballs and mesons should not change as one varies
$\lambda/L$, although the details of the spectrum may change. The limit
$\lambda\gg L$ can be studied using the supergravity description \fiveb.
This was done in \refs{\SakaiCN,\SakaiYT} and we will not review the details
here.

The only point we would like to mention is that unlike the limit $L/R_4\to 0$
discussed in section 4, in the smooth spacetime \fiveb\ a configuration with
separate $D8$ and $\bar{D8}$-branes terminating at the origin does not make
sense, and only the ``hairpin'' shape joining the asymptotic $D8$ and
$\bar{D8}$-branes is possible. Thus, in contrast to the discussion of section
4, where chiral symmetry breaking was simply favored energetically, here it
is mandatory. This is perhaps not surprising given that the `t Hooft matching
conditions {\it require} chiral symmetry breaking for specific values of
$N_f$ \tHooftBH.

The discussion of this section leads us to one of the main points of this
paper. In the case without quarks studied in \WittenZW, the dynamics
depends only on one dimensionless parameter, $\lambda/R_4$. For small
values of this parameter one finds pure Yang-Mills QCD, which is hard to treat
analytically, while for large values one finds a system that differs from
QCD but can be analyzed using supergravity.

In the $D4-D8-\bar{D8}$ system, which contains massless quarks, there
are two dimensionless parameters, $\lambda/L$ and $L/R_4$. For $L/R_4\sim 1$,
the case discussed in \refs{\SakaiCN,\SakaiYT}, the situation is as in
\WittenZW, but for small $L/R_4$ one gets a theory which is
solvable both at weak coupling $\lambda\ll L$ and at strong coupling
$\lambda\gg L$, by using the non-local NJL model of section 3 and the
supergravity analysis of section 4, respectively. Moreover, we presented
evidence that the system is in the same universality class (or phase) for
all values of these parameters. Thus, we see that the brane construction
interpolates between a regime in which the dynamics of the quarks is
described by the non-local NJL model, and one where it is described by
QCD. This might help explain why the NJL model is useful in studying
mesons in QCD \refs{\KlevanskyQE\HatsudaPI\BuballaQV\VolkovKW-\OsipovJS}.

\newsec{Discussion}

Much of the early work  describing large $N_c$ QCD using string theory 
has focused on pure Yang-Mills theory, without quarks. In such theories 
one can compute the spectrum of glueball states, but the results must then 
be compared to lattice QCD since glueballs have not been unambiguously 
identified experimentally.  In trying to obtain more realistic models of QCD 
it is important to incorporate quarks and their meson bound states. The 
$D4-D8-\bar{D8}$ model studied in \refs{\SakaiCN,\SakaiYT} and in this 
paper is  a natural construction  which incorporates the chiral symmetry of 
QCD with massless flavors, describes the spontaneous breaking of this
 symmetry, and leads to a spectrum of meson bound states. 

Previous work on this model \refs{\SakaiCN,\SakaiYT} focused on the
region in the parameter space of brane configurations in which the
separation between the $D8$-branes, $L$, is of the order of the radius
of the extra dimension along the $D4$-branes, $R_4$. If the five-dimensional
`t Hooft coupling $\lambda$ is small, $\lambda\ll L,R_4$, this brane
configuration describes QCD, and is difficult to analyze. For large
$\lambda$ the dynamics is not quite that of QCD but it can be analyzed
using supergravity. There are reasons to believe that the theories one
gets in the two limits are in the same universality class.

In this paper we studied the brane configuration of
\refs{\SakaiCN,\SakaiYT} in the opposite limit $R_4\gg L$.
Consideration of this limit and the space of brane configurations
in general leads to some new insights into the dynamics of quarks
and mesons.

One of the surprising results of our analysis is that there exist
four-dimensional models of quarks without dynamical gauge fields
which exhibit non-trivial infrared dynamics. In local quantum field
theory this is believed to be impossible. Our model contains a
non-local interaction between the left and right-handed quarks, but
since this interaction arises in a D-brane system, we are assured
that it does not lead to any pathologies.

The model in question is a non-local analog of the Nambu-Jona-Lasinio
model, which exhibits dynamical symmetry breaking for arbitrarily
small value of the coupling. While we obtained this model from string 
theory, one can construct it directly in field theory as follows. 
Consider a five-dimensional $U(N_c)$ gauge theory with two codimension 
one defects separated by a distance $L$. At the two defects there are 
$N_f$ left and right-handed fermions in the fundamental representation 
of the gauge group, $q_L$ and $q_R$. The dynamics of the four-dimensional 
fermions and five-dimensional gauge field is governed by the action 
\llfour. The model has a $U(N_f)_L\times U(N_f)_R$ global symmetry. 
Moreover, since the $U(N_c)$ gauge fields are higher dimensional, 
$U(N_c)$ acts as a global symmetry as well.

The dynamical degrees of freedom in the four-dimensional theory are the
fermions $q_L$ and $q_R$. The five-dimensional gauge field provides
an effective coupling between them. This coupling becomes weaker as the
distance $L$ increases, since the five dimensional gauge theory is 
infrared free. In the analogy to superconductivity that motivated 
\NambuTP, $q_L$ and $q_R$ are analogous to the electrons, and the 
five-dimensional $U(N_c)$ gauge bosons are analogous to the phonons.

To study the interaction among the fermions it is convenient to
integrate out the five-dimensional gauge field. In the limit
$L\gg \lambda$, one can do that in the leading, single gluon
exchange approximation. This gives rise to the non-local NJL 
model \lreff, which breaks chiral symmetry  for arbitrarily 
large $L$ (\ie\ arbitrarily weak coupling).

Since the non-local NJL model \lreff\ contains only vector degrees 
of freedom of $U(N_c)$, it is exactly solvable in this limit. It can be 
thought of as a generalization to four dimensions of models such as 
the Gross-Neveu model \GrossJV, and the `t Hooft model \tHooftHX\ 
of two-dimensional QCD.

Compactifying $x^4$ on a circle of radius $R_4$ provides a continuous 
interpolation between the NJL model \lreff\ and QCD. The latter is
obtained in the limit $R_4\simeq L\gg\lambda$. Our analysis of the NJL
model strongly suggests that it is in the same universality class as QCD.

Physically, the reason that the above construction simplifies the study
of quark dynamics is the following. The basic mechanism for chiral 
symmetry breaking and the binding of quarks into mesons is the attractive
interaction between the quarks due to exchange of $U(N_c)$ gauge bosons.
Our model allows one to separate the two scales associated with chiral
symmetry breaking and confinement. In the NJL limit, the mass scale 
associated with chiral symmetry breaking is much higher than that of 
confinement, and all the complications associated with the latter 
disappear in studying the former. In the QCD limit the two scales are 
comparable, which makes the model harder to analyze.

Many interesting questions remain to be addressed. As in previous
discussions, we have dealt with the theory in the limit of zero bare
quark mass. Since one can clearly make sense of the quark mass
perturbation in the NJL model, we expect the same to be true
on the supergravity side. It would be interesting to identify this
perturbation and to study the theory with massive flavors in the
different regimes. This is likely to involve a better understanding
of the $8- \bar 8$ string discussed in section 4. The detailed 
correspondence between bulk and boundary fields also remains to be 
worked out. 

The NJL model has been widely used to study a host of phenomenological 
issues in QCD which are not amenable to perturbation theory, including 
the behavior of the theory at finite temperature and quark density. It 
would be interesting to address these issues in the non-local (gauged) 
NJL model that arises here as well as in its string theory dual.

\bigskip\medskip\noindent
{\bf Acknowledgements:}
We thank O. Aharony, A. Giveon, M. Karliner, O. Lunin, R. Myers, Y. Nambu, 
A. Parnachev, E. Rabinovici,  A. Schwimmer, J. Sonnenschein and 
S. Yankielowicz for discussions. 
DK thanks the Weizmann Institute for hospitality during part of this 
work. The work of EA, JH and SJ  was supported in part by NSF Grant 
No. PHY-0204608. The work of DK was supported in part by DOE grant 
DE-FG02-90ER40560.

\appendix{A}{Conventions and useful results}

We use a ``mainly minus" metric convention. In particular, the four-dimensional flat space
metric is $\eta_{\mu \nu} = {\rm diag}(1,-1,-1,-1)$.  We use the same conventions as
\PeskinEV\ for the Weyl representation of gamma matrices in four dimensions. Thus
\eqn\gamdef{\gamma^\mu = \pmatrix{0 & \sigma^\mu \cr
                                                                    \bar \sigma^\mu & 0 \cr} }
with $\sigma^\mu = (1,\vec \sigma)$ and $\bar \sigma^\mu = (1, - \vec \sigma)$.

We require two Fierz identities which rely on the two relations
\eqn\fiercea{\eqalign{(\sigma^\mu)_{\alphabar \alpha}(\sigma_\mu)_{\betabar \beta} & =
2 \epsilon_{\alphabar \betabar} \epsilon_{\alpha \beta} \cr
				(\sigmabar^\mu)_{\alpha \alphabar} (\sigma_\mu)_{\betabar \beta} & =
2 \delta_{\alpha \beta} \delta_{\alphabar \betabar} \cr }}
the first of which is eqn (3.77) of \PeskinEV\  (which also holds with $\sigma^\mu$ replaced
by $\sigmabar^\mu)$ and the second follows from the first using eqn (3.80)
of \PeskinEV.

Using these, we have for Grassman-valued Weyl spinors
\eqn\fierceb{\left(\psi_{1L}^\dagger \sigmabar^\mu \psi_{2L} \right) \left( \psi_{3R}^\dagger \sigma_\mu \psi_{4R} \right) =
-2 \left( \psi_{1L}^\dagger \psi_{4R} \right) \left( \psi_{3R}^\dagger \psi_{2L} \right) }
and
\eqn\fiercec{\left( \psi_{1L}^\dagger \sigmabar^\mu \psi_{2L} \right) \left( \psi_{3L}^\dagger \sigmabar_{\mu} \psi_{4L} \right) = \left( \psi_{1L}^\dagger \sigmabar^\mu \psi_{4L} \right)
\left( \psi_{3L}^\dagger \sigmabar_{\mu} \psi_{2L} \right) ~. }

Fourier transforms in $d$ spacetime dimensions are given by
\eqn\onea{\eqalign{ \tilde f(k) & = \int d^dx e^{i k\cdot x} f(x)~, \cr
                                     f(x) & = \int {d^dk \over (2
\pi)^d} e^{-ik\cdot x} \tilde f(k) ~. \cr }}
Five-dimensional Fourier transforms of spherically symmetric functions of $k$
in Euclidean space,
\eqn\fivedft{F(x) = \int {d^5k \over (2 \pi)^5} {\tilde F}(|k|) e^{ i k \cdot x} ~,}
can be computed using spherical coordinates to give
\eqn\fivec{F(x) = {1 \over 4 \pi^3} \int_0^\infty dk k^4 {\tilde F}(k) \left( {\sin kx \over (kx)^3} - {\cos kx \over (kx)^2} \right)~.}
As an example, if ${\tilde F}(k) = 1/k^2$ we find the coordinate space propagator
\eqn\fived{F(x) = {1 \over 8 \pi^2} {1 \over (x^2)^{3/2} } ~.}

\listrefs
\end